\documentclass[sn-basic]{sn-jnl}

\jyear{2022}%

\theoremstyle{thmstyleone}%
\theoremstyle{thmstyletwo}%

\theoremstyle{thmstylethree}%

\newcommand{\del}{\mbox{\boldmath $\nabla$}}
\def\vec{\boldsymbol}

\graphicspath{{./fig/}{./png/}}
\usepackage{color}
\usepackage[normalem]{ulem}

\def\bl{Babcock--Leighton}

\def\mm{Maunder minimum}

\newcommand{\Fig}[1]{Figure~\ref{#1}}

\newcommand{\Eq}[1]{Equation~(\ref{#1})}

\begin{document}

\title[Long-term modulation of solar cycles]{Long-term modulation of solar cycles}


\author[1]{\fnm{Akash} \sur{Biswas}}
\author[1]{\fnm{Bidya Binay} \sur{Karak}}
\author*[2]{\fnm{Ilya} \sur{Usoskin}}\email{ilya.usoskin@oulu.fi}
\author[3]{\fnm{Eckhard} \sur{Weisshaar}}

\affil[1]{\orgdiv{Department of Physics, \orgname{Indian Institute of Technology (Banaras Hindu University}, Varanasi}, \orgaddress{
\postcode{221005}, \state{UP}, \country{India}}}
\affil[2]{\orgdiv{Space Physics and Astronomy Research Unit and Sodankyl\"a Geophysical Observatory}, \orgname{University of Oulu}, \city{Oulu}, \postcode{90014}, \country{Finland}}
\affil[3]{\orgdiv{Company}, \orgname{Software \& Automation}, \orgaddress{\street{Brunnenstr. 58}, \city{Bad Nauheim}, \postcode{61231}, \country{Germany}}}


\abstract{
Solar activity has a cyclic nature with the $\approx$11-year Schwabe
cycle dominating its variability on the interannual timescale.
{However, solar cycles are significantly modulated in  length, shape and magnitude, from near-spotless grand minima to very active grand maxima.}
The $\approx$400-year-long direct sunspot-number series is inhomogeneous in quality and too short to study robust parameters of long-term solar variability.
The cosmogenic-isotope proxy extends the timescale to twelve millennia and provides crucial observational constraints of the long-term solar dynamo modulation.
Here, we present a brief up-to-date overview of the long-term variability of solar activity at centennial\,--\,millennial timescales.
The occurrence of grand minima and maxima is discussed as well as the existing quasi-periodicities such as centennial Gleissberg, 201-year Suess/de Vries and 2400-year {Hallstatt} cycles.
It is shown that the solar cycles contain an important random component and have no clock-like phase locking implying a lack of long-term memory.
A brief yet comprehensive review of the theoretical perspectives to explain the observed features in the framework of the dynamo models is presented, including the nonlinearity and stochastic fluctuations in the dynamo.
We keep gaining knowledge of the processes driving solar variability with the new data acquainted and new models developed.
}




\maketitle

\section{Introduction}
\label{sec:Intro}

Sun is a magnetically active star whose activity is a result of the magnetic dynamo process operating in the Sun's convection zone \citep[see, e.g.,][]{Kar14a, Cha20}.
Solar surface magnetic activity varies cyclicly with the main period of about 11 years (called the Schwabe cycle) or, considering inversion of the sign of its magnetic polarity, the 22-year Hale cycle.
More details can be found in an extensive review by \citet{hathawayLR}.
The physics of the dynamo mechanism is currently believed to be {reasonably well understood}.
However, solar cyclicity is far from being a regularly ticking clock and experiences essential long-term variability at timescales longer than the Schwabe cycle.
The solar cycles are not perfectly regular and vary in length, shape, and strength/intensity, or even can enter periods of almost inactive state, called grand minima of solar activity \citep[e.g.,][]{usoskin_LR_17}.

The standard index {quantifying} solar activity is related to sunspot numbers which are available from 1610 AD onward with the quality degrading backwards in time, as discussed in Section~\ref{Sec:SN}.
On one hand, this 410-year-long series exhibits a great deal of variability covering the range from an almost spotless period of the Maunder minimum between 1645\,--\,1715 AD \citep{eddy76} to an epoch of very active Sun between 1940\,--\,2009 called the Modern grand maximum \citep[][]{solanki_Nat_04,usoskin_AA_07}.
This great variability raises important questions, answers to which can put crucial observational constraints on the solar/stellar dynamo theory:
\begin{itemize}
\item
Do the changes between the Maunder minimum and the Modern grand maximum cover the full possible range of solar variability?
\item
Do the grand minima and maxima represent special states of the solar dynamo or simply {represent} the tails of the distribution?
\item How typical are these changes?
\item Do the grand minima episodes appear periodically or randomly?
\item What physical processes drive such changes?
\end{itemize}
The four-century-long sunspot number series is not sufficiently long to answer these questions, and a much longer dataset is needed to form a basis for the answers.
Fortunately, solar activity can be reliably reconstructed from indirect natural proxy data (cosmogenic radioisotopes) on the timescale of 10\,--\,12 millennia, during the period of the Holocene with a stable warm climate on Earth, as discussed in Section~\ref{sec:rec}.
This reconstruction extends the solar-activity dataset by a factor of about 25 making it possible to perform a thorough statistical analysis of solar variability as discussed in Section~\ref{sec:long}, while statistical properties of the solar-cycle modulation are summarized in Section~\ref{sec:stat}.
In Section~\ref{sec:dynamo}, we discuss the implications of the long-term solar variability for the solar dynamo theory and our present level of understanding of the related physics.

\section{Direct Sunspot number series since 1610}
\label{Sec:SN}

Sunspots have been more or less systematically studied since 1610, soon after the invention of the telescope.
Thousands of observational records and drawings exist in archives as being continuously recovered and analyzed \citep[e.g.,][]{vaquero09,arltLR}.
The most recent and continuously updated database of raw sunspot-group observation is collected at the HASO \citep[Historical Archive of Sunspot Observations, http://haso.unex.es/haso --][]{vaquero16}.

Despite numerous observational records, it was noticed only in the middle of the 18th century by the Danish astronomer Christian Horrebow and finally confirmed in the early 19th century by the German observer Heinrich Schwabe, that the number of sunspots varies cyclicly with about
10-year period.
This cycle was later shown to be of about 11 years mean length and appears to be a fundamental feature of solar activity and is now called the \textit{Schwabe} cycle.
{More details of the sunspot number measurements and reconstructions can be found elsewhere in this volume or in comprehensive reviews by \citet{hathawayLR} and \citet{usoskin_LR_17}.}

\subsection{Wolf sunspot series $R_{\rm W}$ and International sunspot number $R_{\rm I}$}

\begin{figure}[t!]
    \centerline{\includegraphics[width=0.8\textwidth]{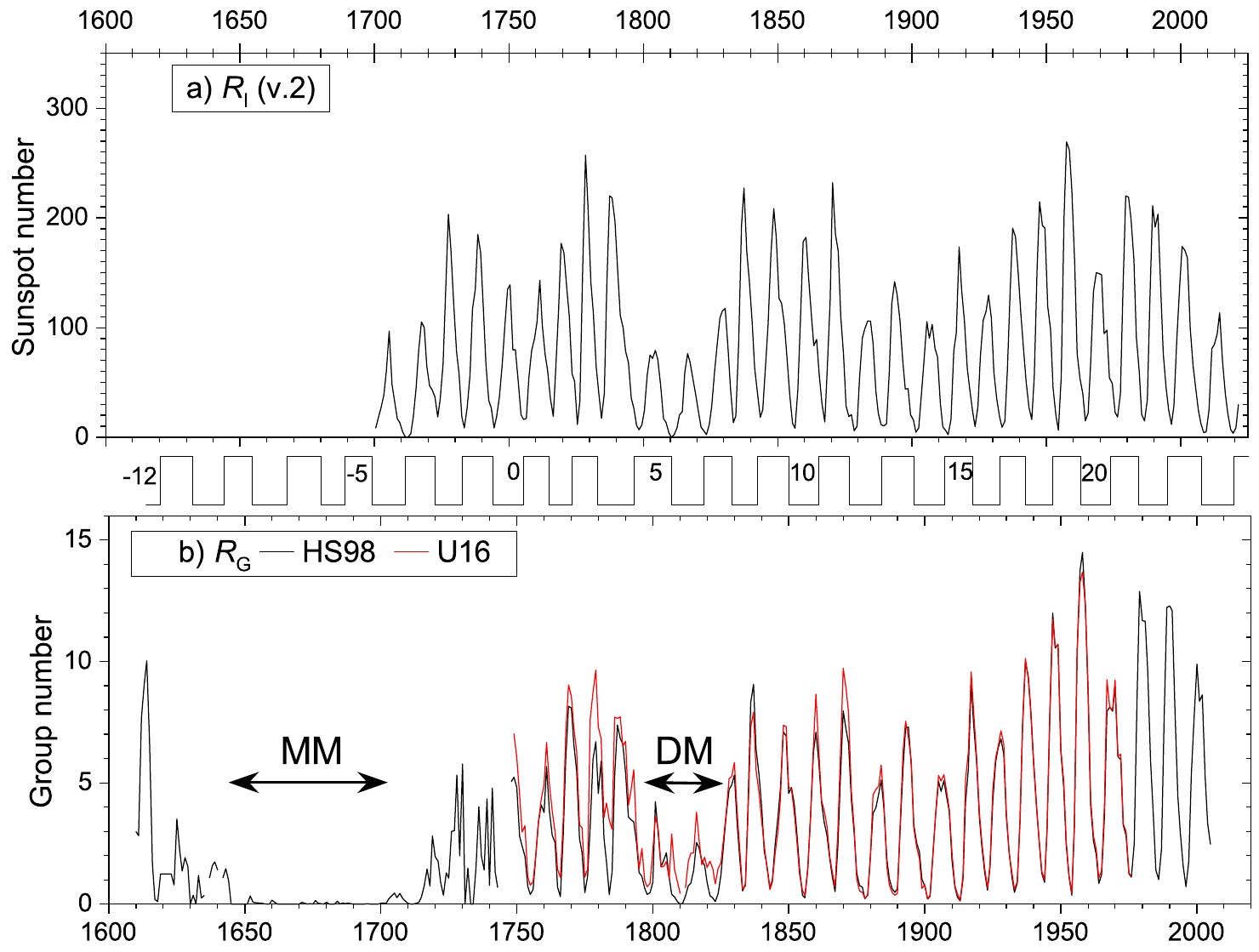}}
    \caption{Annual sunspot activity for the last centuries based on direct sunspot observations:
    a) International sunspot number series version 2 from SILSO (\url{http://sidc.be/silso/dataﬁles}).
    b) Number of sunspot groups according to \citet[][-- HS98]{hoyt98} and \citet[][-- U16]{usoskin_ADF_16}.
    Approximate dates of the Maunder minimum (MM) and Dalton minimum (DM) are shown in the lower panel.
    Standard (Z\"urich) cycle numbering is shown between the panels. {Cycles during the MM are only indicative as provided by \citet{usoskin_MM_AA_00}.}    }
    \label{Fig:SN}
\end{figure}

Following the discovery of the solar cycle, Rudolf Wolf from Z\"urich Observatory founded a synthetic index
{called the sunspot number presently known as Wolf or Z\"urich sunspot number $R_{\rm W}$ (WSN) defined as}
\begin{equation}
    R_{\rm W}= k\cdot (10\cdot G + S),
\label{Eq:Rw}
\end{equation}
where $G$ and $S$ are the numbers of sunspot groups and {all sunspots, including those in groups},  respectively, visible on the solar disc during a given day by the primary observer whose quality scaling factor $k$ is set to reduce his/her counts to the reference observer with $k\equiv$1.
Obviously, the sunspot number is not the same as the number of spots, and for a single sunspot, $R_{\rm W}$=11 assuming $k$=1.
This series, constructed by R. Wolf in 1861 using his own and recovered earlier observations, formally covered the period since 1749 (solar cycle SC \#1 in Wolf's numbering), but in fact, it was more or less reliable only since the 1820's when H. Schwabe started his observations.
Later it was extended back to 1700 with unreliable data.
The compilation of the $R_{\rm W}$ was continued at Z\"urich by Wolf's successors Wolfer, Brunner, Waldmeier and Koeckelenbergh until 1981 when the formation of the sunspot series was  transferred to the Royal Observatory of Belgium \citep{clette07}.

Until 1981, the $R_{\rm W}$ was constructed considering the observation of only one primary observer for each day, all other observations were discarded.
This series could not, till now, be revisited or redone because of the lack of original raw data.
Accordingly, when several {apparent inhomogeneities were found in the standard Wolf sunspot} series \citep{leussu13,clette14,lockwood_1_14}, only step-wise corrections to the old series could be done \citep{clette14,clette16}.
This `corrected' sunspot series is known as the International sunspot number series version 2.0, $R_{\rm I}$(2.0), and is available at the SILSO (Sunspot Index and Long-term Solar Observations, \url{https://www.sidc.be/silso/datafiles}) formally since 1700.
The $R_{\rm I}$(2.0) is shown in Figure~\ref{Fig:SN}a along with the standard Z\"urich sunspot cycle numbering.

Although the update of the series was through several adjustments of scaling jumps, an important effort is currently done by the community to restore and digitize old raw data \citep{clette21} so that it will be possible to redo the sunspot number series from scratch increasing its reliability and assessing realistic uncertainties.

\subsection{Group sunspot number series $G_{\rm N}$}

Since the sunspot number (Equation~\ref{Eq:Rw}) includes both numbers of sunspot groups (weighted by a factor of 10) and individual sunspots, it is sensitive to the quality of observations.
This was addressed by \citet{hoyt98} who noticed that sunspot groups are defined more reliably than individual spots and created the group sunspot number series $G_{\rm N}$ which is simply the number of sunspot groups $G$ on the solar disc corrected for the observer's quality.
This series is shown in Figure~\ref{Fig:SN}b.
Sometimes it is scaled up to match the values typical for $R_{\rm W}$.
However, contrary to $R_{\rm W}$, $G_{\rm N}$ is based on the average of all available observations for each day, not only the primary ones.
Another principal difference between $R_{\rm W}$ and $G_{\rm N}$ is that \citet{hoyt98} created and published a full database of raw data they used to construct the $G_{\rm N}$ series.
Accordingly, this series can be completely redone as a whole, without limitation to the `correction factors'.

It was recognized that the original $G_{\rm N}$ underestimated solar activity during the 19-th century \citep{clette14}, and several efforts have been made to revisit it using different methodologies and inter-calibrations \citep[e.g.,][]{svalgaard16,usoskin_ADF_16,chatzistergos17,willamo17}.
One of the reconstructions is also shown in Figure~\ref{Fig:SN}b.
However, these new series often moderately disagree with each other illustrating the problem of compiling a homogeneous series from individual raw datasets \citep{munos19}.
It is presently impossible to {decide} between different reconstructions of the group sunspot series, but the zoo of those gives a clue of what the related uncertainties are, and presently they are bounded by the series of \citet{svalgaard16} from the top and  from below by \citet{hoyt98}.

\section{Cosmogenic-isotope-based reconstructions of long-term solar variability}
\label{sec:rec}

The sunspot number series covers ca. 410 years in the past with the quality degrading back in time \citep{munos19} and principally cannot be extended before the 17-th century because of the lack of instrumental data.
Unaided (naked-eye) observations of sunspots do not provide systematic quantitative information on solar activity \citep{usoskin_LR_17}.
There are some other proxy-based indices of solar activity, such as geomagnetic or heliospheric activity, and radio-emission of the Sun, but they all are based on scientific measurements and typically do not go beyond the middle of the 19-th century.
Fortunately, there is one solar-activity proxy which can help in reconstructing solar variability on the multi-millennial timescale.
This is related to cosmogenic radioisotopes which are produced and preserved in dateable archives in a natural way.

\subsection{Method of cosmogenic isotopes}

Solar surface magnetic activity and hot corona create the solar wind which is a  supersonic outflow of solar coronal plasma permanently emitted from the Sun \citep[see, e.g.,][]{vidotto_LR}.
Because of its high conductivity, solar wind drags away the solar magnetic field which appears `frozen' in the solar-wind plasma.
This wind radially expands forming the heliosphere, a region of about 200 astronomical units across which is totally controlled (in the magnetohydrodynamical sense) by the solar wind and magnetic field \citep[e.g.,][]{owens13}.
The heliosphere makes an obstacle for charged highly energetic particles of galactic cosmic rays (GCRs) which permanently bombard it isotropically with nearly constant flux.
Inside the heliosphere, cosmic rays are affected by four major processes, viz. scattering and diffusion on magnetic irregularities, convection by expanding solar wind, adiabatic cooling, and large-scale drifts.
All these processes are ultimately driven by solar activity leading to the solar modulation of cosmic-ray flux near Earth so that the cosmic-ray flux is stronger when solar activity is weak and vice-versa \citep[e.g.,][]{potgieterLR}.
Thus, knowing the modulated flux of GCRs at a moment in time, one can assess the level of solar activity  {slightly before that \citep[within one year -- ][]{koldobskiy_SP_22}}.
Of course, there were no scientific cosmic-ray detectors in the distant past, but there is a natural cosmic-ray monitor -- cosmogenic radioisotopes.

Cosmogenic radioisotopes are unstable nuclides, which cannot survive from the time of the solar-system formation, and whose main source is related to nuclear reactions caused by cosmic rays in the Earth's atmosphere \citep{beer12}.
After production in the atmosphere by GCR, nuclides can be stored in natural independently dateable archives, such as tree trunks, polar ice cores, lake/marine sediments, etc.
Accordingly, the flux of GCR can be estimated in the past by measuring the abundance of such isotopes in the archives, forming the only quantitative proxy of solar activity over long timescales \citep[see more details in][]{beer_NM_00,usoskin_LR_17}.
The most important cosmogenic isotopes are $^{14}$C `radiocarbon' (half-life 5730 years) measured in dendrochronologically dated tree rings and $^{10}$Be ($\approx 1.4\cdot 10^6$ years) measured in glaciologically dated ice cores.

Conversion between the measured isotope concentration and production by cosmic rays requires a knowledge of the isotope's transport and deposition processes which are currently well modelled \citep[e.g.,][]{roth13,heikkila13,golubenko21}.
Additionally, it needs to be corrected for the changing geomagnetic field \citep[e.g.,][]{pavon18}, and the resulting variability can be attributed to solar activity.
The conversion from the cosmic-ray modulation to the heliospheric properties
({open} solar flux) and then to the pseudo-sunspot numbers is done via a chain of physics-based models making it possible to reconstruct solar activity and the related uncertainties \citep[see, e.g.,][]{usoskin_LR_17,wu18}.

\subsection{Holocene ($\approx$12 kyr) decadal reconstruction}

\begin{figure}[t!]
    \centerline{\includegraphics[width=0.9\textwidth]{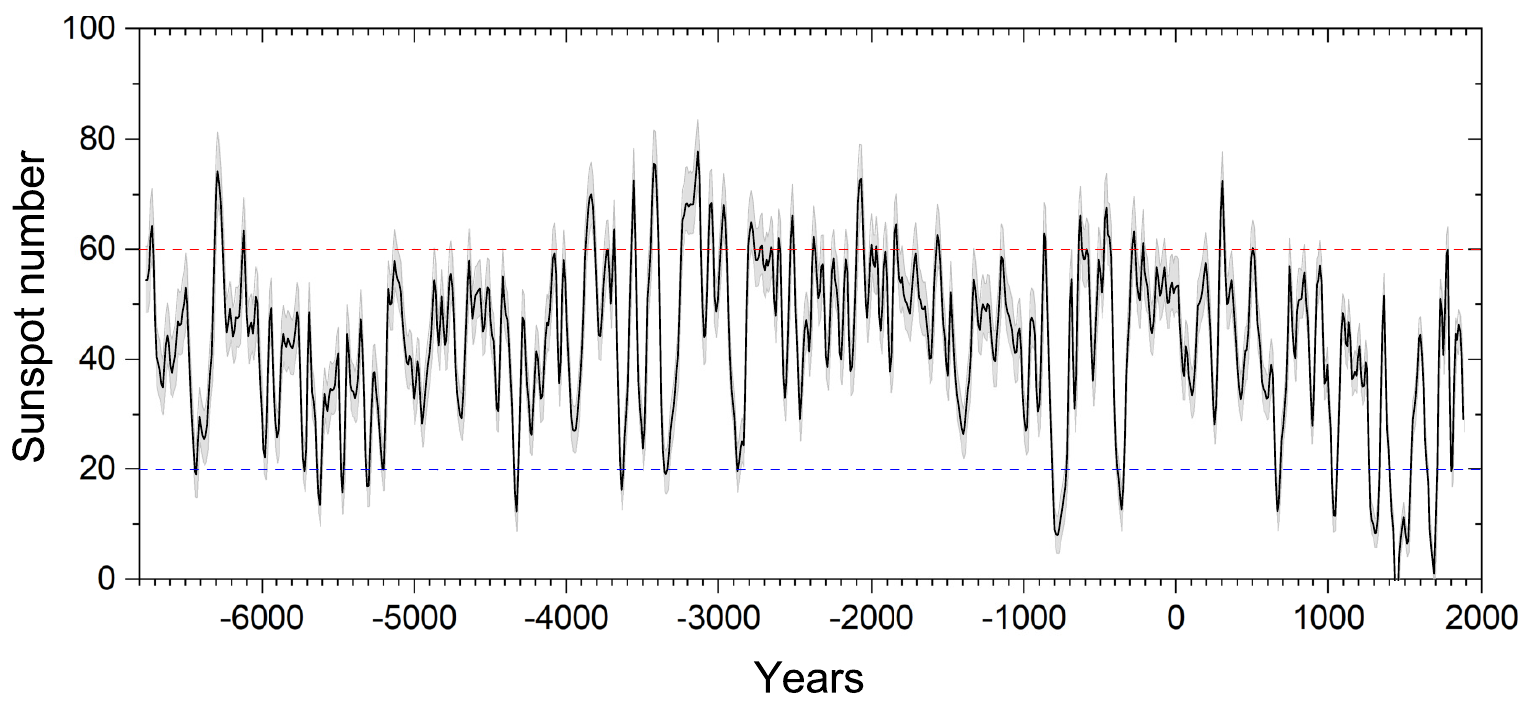}}
    \caption{Multi-proxy reconstruction of the decadal sunspot numbers (in the classical Wolf's definition) over the last nine millennia, along with the $1\sigma$ uncertainties \citep{wu18}.
     {The blue and red dashed lines approximately denote the low (Grand minimum) and high states of solar activity.}}
    \label{Fig:Wu}
\end{figure}

While the idea of the use of cosmogenic-isotope data as a proxy to solar activity has been discussed since long \citep{stuiver61,lal62}, first approaches were empirical as based on timescale separation of the cosmogenic data: timescales longer than 500 years were thought to be caused by changes in the large-scale geomagnetic field, while shorter time scales -- by solar activity \citep{damon91}.
That approach made it possible to identify grand solar minima \citep{eddy76,stuiver89} but was unable to provide a quantitative reconstruction of solar activity because both factors are important at the centennial timescales.
A full reconstruction of solar activity from cosmogenic-isotope data became possible only after the development of models of cosmic-ray-induced atmospheric cascades \citep{masarik99}.
The first quantitative reconstruction of solar activity using a physics-based approach was made by \citet{usoskin_PRL_03} on the millennial time scale \citep[see also][]{solanki_Nat_04}.
Later the reconstructions were extended to the Holocene (the present period of stable warm climate lasting for about 12 millennia) using different cosmogenic isotopes \citep[e.g.,][]{vonmoos06,steinhilber12,usoskin_ADF_16}.
The most recent and accurate multi-millennial solar-activity reconstruction by \citet{wu18} is based on a multi-proxy Bayesian approach providing also realistic uncertainties.
It is shown in Figure~\ref{Fig:Wu}.
One can see that solar activity varies essentially between the grand minima, visible at sharp dips down to 10\,--\,20 (in sunspot number, SN), and grand maxima when SN exceeds 60, while most of the time the solar-activity level remains moderate at SN$=40\pm 10$ \citep[see more detail in][]{usoskin_AAL_14}.
The results of an analysis of the solar-activity variability are reviewed in Section~\ref{sec:long}.

Because of the low time resolution of the cosmogenic-isotope throughout the Holocene \citep[typically decadal -- see, e.g.,][]{reimer20}, reconstructions of solar activity are also usually limited to the 10-year resolution being thus unable to resolve individual solar cycles.
Long-term reconstructions of solar activity are limited to the Holocene timescale because of the stable climate so that the standard models of the isotope atmospheric transport and deposition can apply.
However, for the ice-age-type of climate, the properties of the atmospheric transport are quite uncertain including the large-scale atmospheric and ocean circulation, which prevents quantitative assessment of solar activity.
At present, there is no model which is
able to handle this in a satisfactory manner, but progress is expected in the future.

\subsection{$\approx$100 solar cycles reconstructed}

\begin{figure}[t!]
    \centerline{\includegraphics[width=0.9\textwidth]{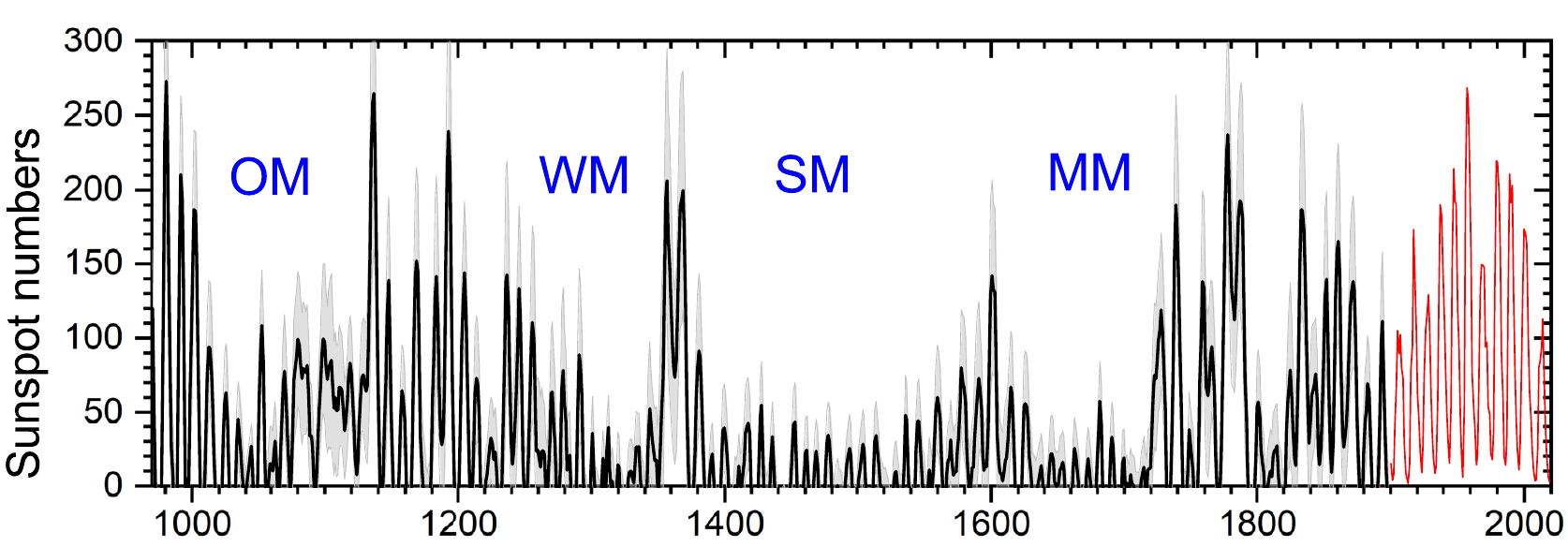}}
    \caption{Annual reconstruction, based on high-precision $^{14}$C data, of the sunspot numbers over the last millennium (970\,--\,1900), along with the $1\sigma$ uncertainties \citep{usoskin_AA_21}.
    The red curve presents the ISN (v.2) since 1900.
    Approximate periods of the Oort (OM), Wolf (WM), Sp\"orer (SM) and Maunder (MM) grand minima are indicated in blue letters.}
    \label{Fig:1000}
\end{figure}

Thanks to the recent technological progress, high-precision measurements of annual $^{14}$C concentrations have been performed with the annual resolution for the last millennium \citep{brehm21}.
It allowed us to make, by applying the physics-based model, the first reliable reconstruction of individual solar cycles beyond the epoch of telescopic observations \citep{usoskin_AA_21} as shown in Figure~\ref{Fig:1000}.
Four known grand minima are seen -- Oort, Wolf, Sp\"orer and Maunder minima, and between the minima, there are clear solar cycles of variable amplitude.
In this way, 85 individual solar cycles have been reconstructed from $^{14}$C of which 35 cycles are reasonably and well resolved, 21 are poorly and 29 are not reliably resolved, mostly during the grand minima of activity.
Overall, including both direct solar observations and proxy-based reconstructions, we now have information on 96 solar cycles of which 50 are well resolved, thus nearly tripling the extent of the solar-cycle knowledge and doubling the number of well-defined cycles.

The extended statistic made it possible to perform a primary analysis of the solar-cycle parameters.
The length of the well-defined cycles was $10.8\pm 1.4$ years which is in good agreement with $11.0\pm 1.1$ years known for the ISN dataset.
The statistical significance of the Waldmeier rule (solar-cycle height is inversely correlated with the length of the ascending phase -- high cycles rise fast) has been confirmed with the extended dataset, implying its robust nature \citep{usoskin_AA_21}.
However, the Gnevyshev-Ohl rule of even--odd cycle pairing \citep{gnevyshev48,usoskin_lost_AA_01} has not been confirmed, nor rejected with the extended data.
A more detailed analysis of this new dataset is still pending.

\section{Long-term solar activity}
\label{sec:long}

With the reconstructed long series, one can investigate properties of solar variability which pose observational constraints crucially important for solar physics but cannot be set by the too short-ranging conventional direct telescopic observations of the Sun.
While the 11-year solar cycle forms the main feature of solar activity, the cycles are far from being perfect clock ticks -- they vary by both duration and amplitude including periods of greatly suppressed activity, grand minima (see Figure~\ref{Fig:1000}).
Here we review the most important features of long-term solar variability.

\subsection{Long quasi-periodic variations (Gleissberg, Suess/de Vries, Hallstatt cycles)}
\label{Ss:cycles}
\begin{figure}[t!]
    \centerline{\includegraphics[width=0.9\textwidth]{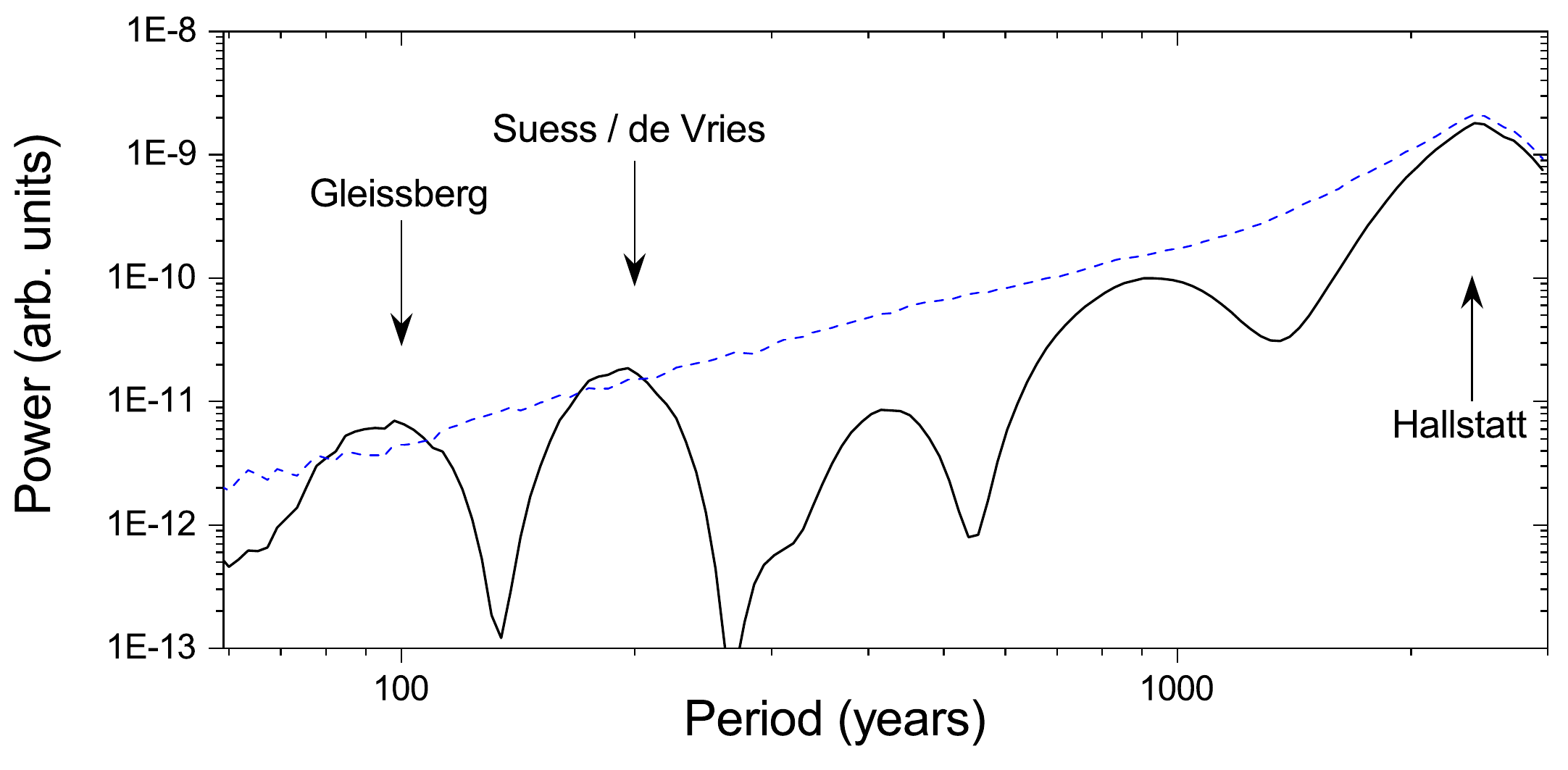}}
    \caption{Global wavelet (Morlet basis) power spectrum (black curve) of the long-term sunspot-number series shown in Figure~\ref{Fig:Wu}.
    Blue-dashed line denotes the 90\% confidence level estimated using the AR1 auto-regressive noise, following the methodology of \citet{grinsted04}.
    {Approximate locations of the discussed quasi-periodic variations (Section~\ref{Ss:cycles}) are indicated by vertical arrows.}
    }
    \label{Fig:wv1000}
\end{figure}

It is hardly possible to distinguish whether solar variability on a long-term scale (Figure~\ref{Fig:Wu}) is stochastic/chaotic or (quasi)periodic.
Power-spectrum analyses are controversial but generally agree that there are three period ranges with {apparent} and barely significant variability.
An example of the global wavelet power spectrum is shown in Figure~\ref{Fig:wv1000}.

One is the centennial variability, called the \textit{Gleissberg} cycle, which is not a strict periodicity but a characteristic period range between 60\,--\,140 years \citep[e.g.,][]{peristykh03,ogurtsov04}.
The Gleissberg cycle is clearly seen in the direct sunspot data but is less pronounced throughout the Holocene.

Another important periodicity is the \textit{Suess} cycle (called also \textit{de Vries} cycle in the literature), which has a narrow period range between 200\,--\,210 years and an intermittent occurrence.
It is typically seen as a recurrence of grand minima within clusters of reduced solar activity \citep{usoskin_AAL_14} as seen, e.g., in Figure~\ref{Fig:1000}, but is not readily observed during the epochs of moderate solar activity.

Sometimes, the so-called \textit{Eddy} millennial cycle is claimed to exist \citep{steinhilber12}, but it is unstable and cannot be identified in a significant way (see Figure~\ref{Fig:wv1000}).

Additionally, there exists a very-long cycle with a timescale of 2000\,--\,2400 years called the \textit{Hallstatt} cycle \citep{damon91,vassiliev02,usoskin_AA_16}.
Because of its length, it cannot be robustly defined in the $\approx$10-kyr time series (see Figure~\ref{Fig:wv1000}).
The nature of the Hallstatt cycle is still unclear: it is likely to be ascribed to the Sun \citep{usoskin_AA_16} but geomagnetic or climatic origin cannot be excluded.
{Longer-scale variability cannot be reliably assessed from the cosmogenic-isotope data, in particular, because of the unresolved discrepancy between $^{14}$C and $^{10}$Be datasets on the multi-millennial timescale as probably related to the effect of deglaciation \citep[e.g.,][]{vonmoos06,usoskin_AA_16,wu18}.}

\subsection{Grand minima and maxima}

As seen, e.g., in Figures~\ref{Fig:Wu} and \ref{Fig:1000}, solar activity sometimes drops fast, within one--two solar cycles, to the very quiet level with almost no sunspots on the solar surface.
These drops are called grand minima of activity.
{Until the 1970s, the existence of such minima was debated}, but \cite{eddy76} had convincingly proved that the sunspot activity {indeed dropped} to almost no sunspots between 1645\,--\,1715 as confirmed also by other proxies such as auroral displays at mid-latitudes.
That grand minimum was called the \textit{Maunder} minimum.
More grand minima have been found later using the cosmogenic-isotope data \citep[e.g.,][]{usoskin_AA_07,inceoglu15}.
At present, about 30 grand minima of duration ranging between 40\,--\,70 (Maunder-type minima) and 100\,--\,140 years (Sp\"orer-type) each, have been identified during the Holocene occupying about 1/6 of the time.
It has been shown that the grand minima correspond to a special state of the solar dynamo \citep[e.g.,][]{usoskin_AAL_14}.

Solar activity was abnormally high in the second half of the 20th century compared to the 19th or 21st centuries \citep{lockwood99} but it was unknown whether this high level is unique or typical.
{Using the cosmogenic-isotope data, it was discovered that the period from the 1940s to 2010 was not unique and there are other similarly high but very rare episodes, that forms the concept of a \textit{grand solar maximum}} \citep{usoskin_PRL_03,solanki_Nat_04}.
Grand maxima represent periods of enhanced solar activity covering at least a few solar cycles.
There were about 20 grand maxima over the Holocene which cover $\approx$10 \% of the time \citep{usoskin_AA_07,inceoglu15}, but they are defined not as robustly as grand minima.
No apparent clustering in the grand-maxima occurrence or duration has been found, nor do they form a special distribution of solar cycles \citep{usoskin_AAL_14,usoskin_AA_16}.
It is still unknown whether grand maxima make a special mode of the dynamo, similar to grand minima, or just represent a rare tail of the solar-cycle-strength distribution.

\section{Statistical properties of the long-term modulation of solar cycles}
\label{sec:stat}

As historical records show, solar cycles are highly variable in amplitude and length.
The validity of theoretical models that attempt to predict this variability depends heavily on whether the cycle exhibits long-term phase stability or whether the phase is subject to a random walk, {or a mixture of these}.
In the first of the two extreme cases, the system has infinite phase memory and in the second case no phase memory at all.
Phase stability could be achieved through synchronization processes, such as high-quality torsional oscillations in the solar interior \citep{Dicke:1970} or the weak tidal forces of planets \citep[e.g.,][]{Stefani:etal:2021}.
Dynamo models generally predict phase progression without memory.
An insightful summary of the use of historical observations to explain solar phenomena was given by \citet{vaquero09}.

{The question of the regularities and randomness of solar activity variability has been studied for a long time.
For example, statistical methods including those based on the Lyapunov and Hurst exponents or Kolmogorov entropy \citep[e.g.,][]{ostryakov90,mundt91,carbonell94,ruzmaikin94,lepretti21} were inconclusive, implying that a mixture of different components is likely \citep[see more details in][]{usoskin_LR_17,Petrovay20}.}

Various publications \citep[e.g.,][]{Lomb:2013,Russell:etal:2019, Stefani:etal:2020} claim that the solar cycle is phase stable.
However, to answer the question of whether the phase is stable or not, one needs a clear definition of phase stability, an appropriate statistical analysis as well as reliable data on which to apply the analysis.
\citet{Dicke:1978} and \citet{Gough:1978} were {among} the first to perform a systematic statistical analysis based on telescopic sunspot records.
Independently, but using similar concepts, they concluded that the time span of the available data was too small for a clear distinction between the two cases.
Later, \citet{Gough:1981,Gough:1983,Gough:1988} corrected and modified his earlier analysis without altering the conclusion.

Interestingly, \citet{Eddington:etal:1929} analyzed the light-curve variations of long-period variable stars, a problem close to the variability of the solar cycle.
By deriving a statistical function to which the processed observational data were fitted, they were able to determine two indicators for the {composition of clock-synchronised phase perturbations and random phase perturbations of the light signal.}

\begin{figure}[t!]
    \centerline{\includegraphics[width=0.8\columnwidth]{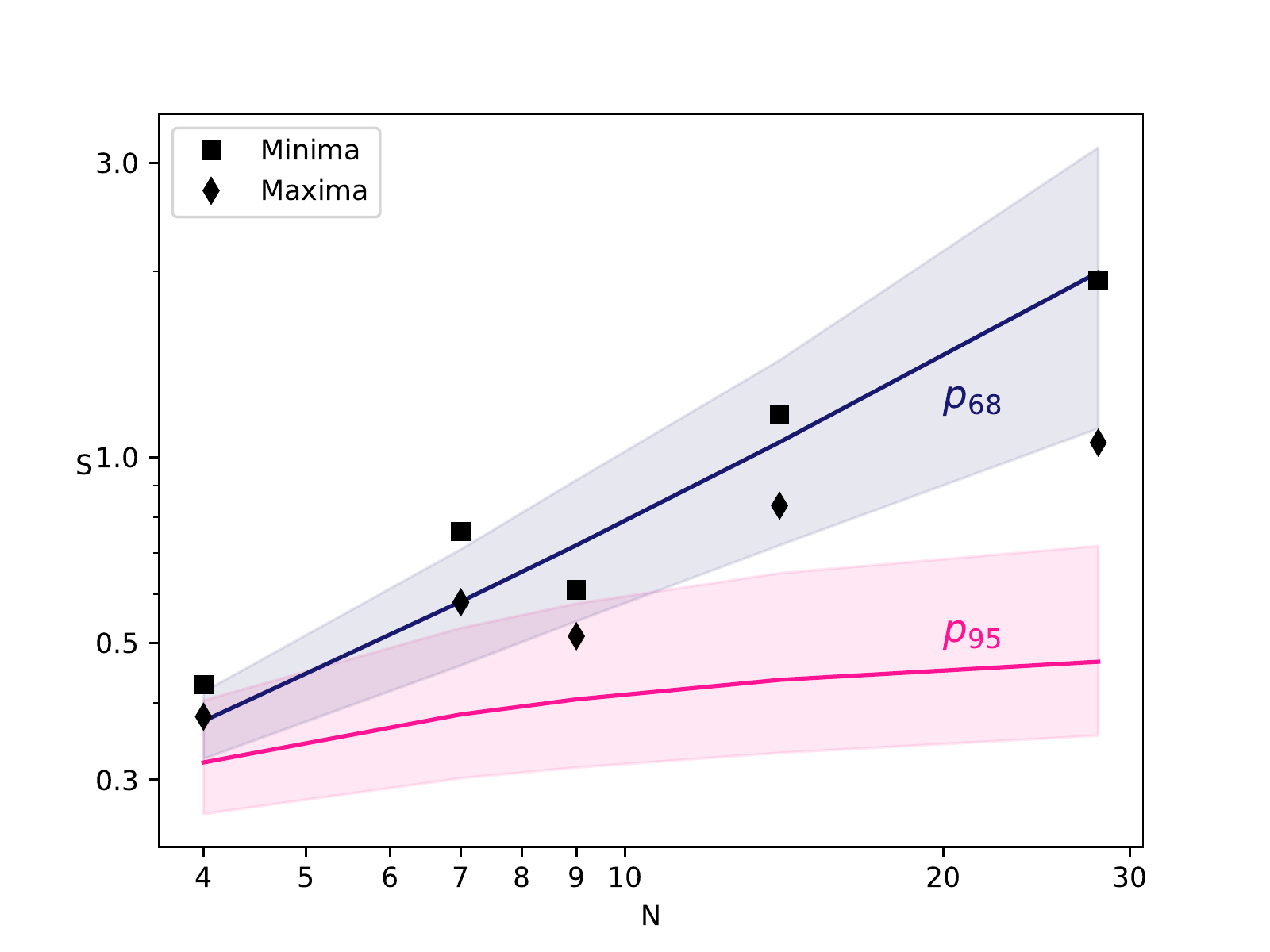}}
    \caption{Modified Gough test $S$ applied to the epochs of sunspot minima and maxima of 28 activity cycles between 1712 and 2019.
    Symbols correspond to the solar cycle maxima and minima, as denoted in the legend.
    The black line with the shaded 68\% confidence interval depicts the random phase hypothesis (Eq.~\ref{Eq:Sr}).
    The {red} curve with the shaded 95\% c.i. depicts the clock phase hypothesis (Eq.~\ref{Eq:Sc}). }
    \label{Fig:28}
\end{figure}
\citet{Weisshaar:etal:2023} have revisited Gough’s analysis based on newly available data.
For clarity, a brief outline of Gough's test is given here:
From the arithmetic mean of the individual cycle lengths \citep{Gough:1981}, the regular minima or maxima of the hypothetical dynamo or clock cycles and thus the corresponding phase deviations can be determined as the difference to the observed minima or maxima.
The basic statistics are the expectation values of the variances of cycle period, $E(\sigma_P^2)$, and phase, $E(\sigma_\phi^2)$.
The final statistics is defined as the ratio of the two variances to cancel out the unknown fluctuation amplitude:
\begin{equation}
S = {E(\sigma{_\phi}^2)\over E(\sigma{_P}^2)}
\end{equation}
Later, \citet{Gough:1983} modified the method by replacing the arithmetic mean of the cycle period with a value that minimizes the variance of the phase deviations, resulting in a more sensitive distinction between the clock regime and the random phase regime.
Calculating the expectation values of the variances for the two cases, one obtains the following expressions for $S_c$ (clock) and $S_r$ (random phase) using the modified method:
\begin{equation}
S_c = {E(\sigma{_\phi}^2)\over E(\sigma{_P}^2)} = {N^2\over 2(N+1)^2}
\label{Eq:Sc}
\end{equation}
which asymptotically reaches  $N \to \infty$, ${S_c} \to {1\over 2}$;
\begin{equation}
S_r = {E(\sigma{_\phi}^2)\over E(\sigma{_P}^2)} = {N(N+3)\over 15(N+1)}
\label{Eq:Sr}
\end{equation}
which asymptotically reaches $N \to \infty$, ${S_r} \to {N\over 15}$.

\begin{figure}[t!]
    \centerline{\includegraphics[width=0.8\columnwidth]{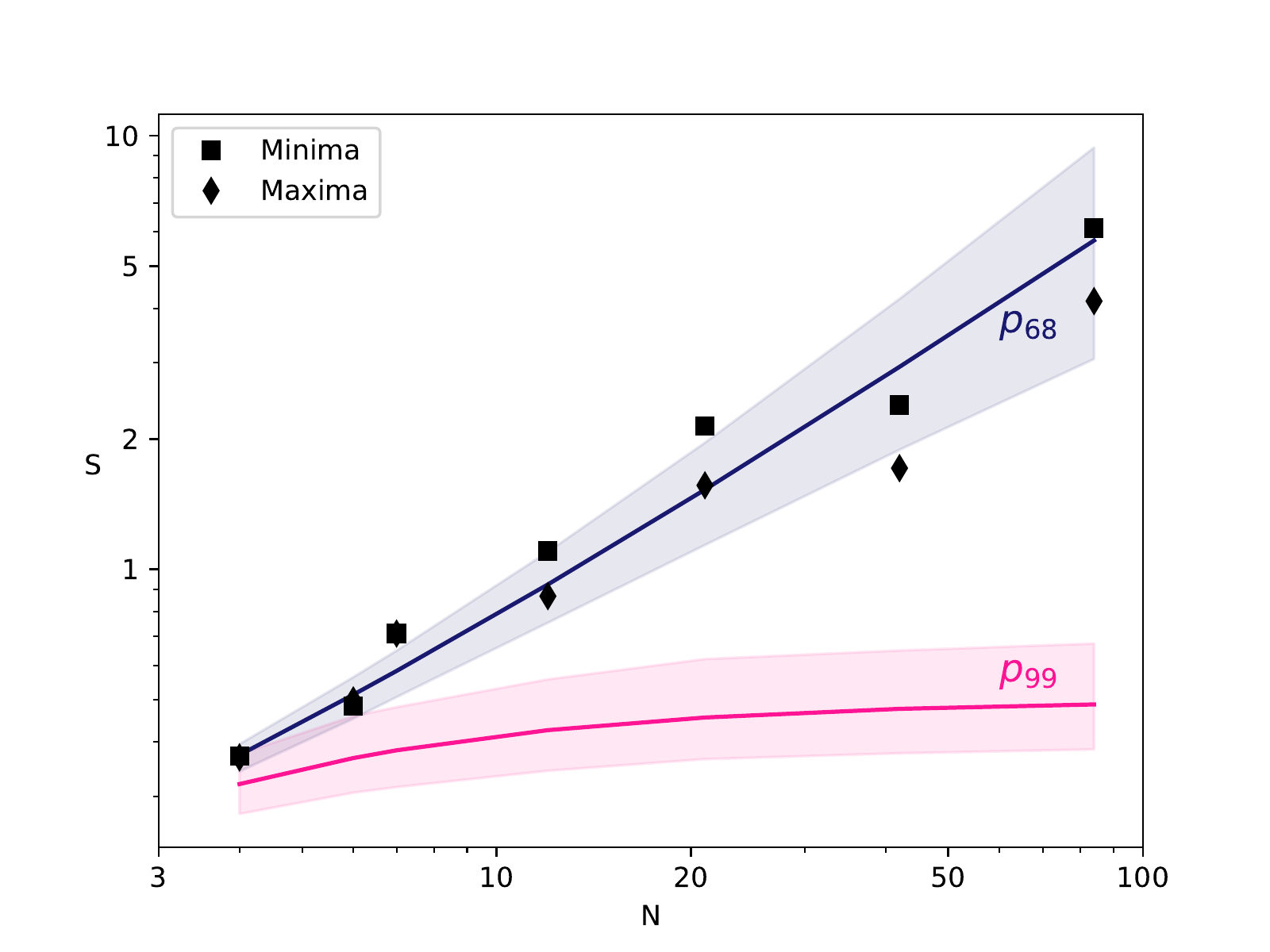}}
    \caption{Modified Gough test (notations are similar to those in Fig.~\ref{Fig:28}) applied to the series of 84 cycles covering the period between 976 and 1999 as reconstructed from $^{14}$C data by \citep{usoskin_AA_21}.
    The data agree with a random phase shift, while synchronization with the "clock" is rejected at the confidence level much higher than 99\% due to the longer data set.}
    \label{Fig:84}
\end{figure}

The procedure to apply Gough’s test to an observed data set is as follows: The data set is divided into contiguous segments of $N$ cycles each.
Then the ratio of the averages of the empirical variances is calculated and compared with the ratio of the expectation values, plotted as functions of $N$ in \Fig{Fig:84}.

\citet{Weisshaar:etal:2023} augmented the method by determining suitable confidence intervals through Monte Carlo simulations for the clock and the random phase cases, assuming normally distributed variations in cycle length.
They applied the test to the extended sunspot record of now 28 cycles, four more than available to Gough.
The main improvement is narrower confidence intervals, rejecting the synchronization hypothesis on a 2$\sigma$ level (Figure~\ref{Fig:28}).

Recently, a reconstruction of yearly sunspot numbers from the record of cosmogenic $^{14}$C in tree rings for the years 976 until 1888 \citep{brehm21, usoskin_AA_21} has extended the number of contiguous cycles available for the analysis to 84.
The Gough test confirms the previous result based on the direct sunspot record, in fact strengthening it significantly, since now the synchronization hypothesis can be rejected even on a $>3\sigma$ level (Figure~\ref{Fig:84}).

{\citet{Weisshaar:etal:2023} also applied the method of \citet{Eddington:etal:1929} mentioned above to these new data and found, consistent with the analysis discussed here, that the fraction of clock-synchronised perturbations is negligible.}

{The question may arise how misidentifications of the observed solar cycles can affect the results.
If this happens not too common, the nature of the fluctuations (phase stability or migration) is not expected to be changed by this bias.
As a test, a lost cycle between more distant minima was ``restored'' by placing a minimum in between.
This did not cause the $S$-values to leave the phase migration confidence interval.}

Furthermore, the above-mentioned method of the phase evolution of empirical cycle data is therefore consistent with a random walk (such as provided by a memory-less dynamo process).
External synchronization by a `clock' is clearly excluded at a high significance.

\section{Implications for the dynamo theory}
\label{sec:dynamo}
The solar magnetic cycle is maintained by a dynamo process, operating in the solar convection zone (SCZ).
Thus, it is natural to expect that the variations in the solar cycle are caused by some mechanisms in the solar dynamo.
Here we identify the causes of the variations in the solar cycle and demonstrate them by presenting results from some illustrative models.
Let us first summarise the mechanism of the solar dynamo.

\subsection{Introduction to the solar dynamo}
There is enough evidence that the solar dynamo is a mechanism in which toroidal and poloidal fields sustain each other through a cyclic loop \citep[e.g.,][]{Pa55, CS15}.
In this loop, the toroidal field is generated due to the shearing of the poloidal field by the differential rotation in the deeper CZ.
The toroidal field rises to the surface due to magnetic buoyancy to give rise to sunspots or more generally bipolar magnetic regions (BMRs).
These BMRs are systematically tilted with respect to their East-West orientations.
Due to these tilts, after their decay,
BMRs produce a poloidal field.
This, the so-called \bl\ process is clearly identified in the observed magnetic field data on the solar surface \citep[e.g.,][]{Mord22}.
The observed correlation between the polar field (or its proxy) at the solar minima
and the amplitude of the next cycle \citep{WS09, KO11, Muno13, Priy14} and the flux budgets of the observed and the generated poloidal and toroidal fields \citep{CS15} suggest that the \bl\ process is possibly the main source of the poloidal field in the Sun.

There is however another mechanism through which the poloidal field in the sun can be produced
and that is the classical $\alpha$ effect as originally proposed by \citet{Pa55} and mathematically formulated by \citet{SKR66}.
In this mechanism, the toroidal field is twisted by the helically rising blobs of plasma in the SCZ.
However, this process of lifting and twisting of the field by the convective flow experiences catastrophic quenching due to helicity conservation
and thus this process operates when the energy density of
the toroidal field is less than the energy density of the convective motion
 \citep[Sec. 8.7 of][]{BS05}.
Therefore, this $\alpha$ effect is unfavourable in the solar convection zone and the obvious option is to consider the observationally supported \bl\ process for the generation of the poloidal field in the sun.

To study the dynamo action, we need to begin with at least following two fundamental equations of magnetohydrodynamics (MHD).
\begin{equation}
 \frac{\partial \vec{B}}{\partial t} =  \vec{\nabla} \times (\vec{v} \times \vec{B} - \eta \vec{\nabla} \times \vec{B}),
 \label{eq:ind}
\end{equation}
\begin{equation}  
  \label{eq:mom}
  \rho \left[\frac{\partial \vec{v}}{\partial t} + (\vec{v} \cdot \del) \vec{v} \right] =  -\del P + \vec{J}\times\vec{B}+ \del \cdot (2 \nu \rho S) + \vec{F},
\end{equation}
where $\vec{B}$ and $\vec{v}$ are the magnetic and velocity fields, respectively, $\eta$ is the magnetic diffusivity, $\rho$ is the density, $P$ is the pressure, $\vec{J} = \del \times \vec{B}/\mu_0$, the current density, $\nu$ is the kinetic viscosity, $S_{ij} = \frac{1}{2} ( \nabla_i v_j + \nabla_j v_i) - \frac{1}{3} \delta_{ij} \del \cdot \vec{v}$ is the rate-of-strain tensor, and the term $\vec{F}$ includes gravitational, Coriolis and any other body forces acting on the fluid.
These equations along with the mass continuity and energy equations and equation of state are numerically solved
with appropriate boundary conditions in the solar CZ to study the dynamo problem.
Broadly there are two approaches for doing this, namely, the global MHD simulations and mean-field modellings.
In global MHD simulations, we solve the above MHD equations
numerically to resolve the full spectrum of turbulent convection.
In mean-field models, we study the evolution of the mean/large-scale quantities by parameterizing the small-scale/fluctuating quantities using suitable approximations.

Global MHD simulations for the Sun are challenging due to extreme parameter regimes, such as high fluid and magnetic Reynolds numbers and large stratification. Despite these, simulations have begun to produce some solar-like features; see Section~6 of \citet{Cha20}.
However, due to their computationally expensive nature, these simulations were rarely run for many cycles so that the cycle variabilities can be studied. \citet{PC14} have produced simulations for several cycles and shown long-term modulations \citep[also see][for a simulation at solar rotation rate although ran for not many cycles]{Kar15}.
\citet{ABMT15} and \citet{Kap16} performed MHD convection simulations for the cases of three and five times the solar rotation rate, respectively.
They both found an episode of suppressed surface activity, somewhat resembling the solar grand minimum.
Although these results of cycle modulations are encouraging,
simulations face serious issues when matching with observations, for example, concerning
solar observations, simulations (i) produce higher power at the largest length scale, (ii) do not produce BMRs,
and (iii) do not produce correct large-scale flows, particularly, they produce a large variation in the differential rotation.

On the other hand, mean-field models are computationally less expensive and easy to analyse their results. Probably due to these reasons, long-term modulations are studied using mean-field dynamo models. Due to the observational facts that the magnetic field at the solar minima and the large-scale velocity field are largely axisymmetric,
historically the mean-field models are constructed under axisymmetric approximation.  With this approximation,
the equations for the poloidal and toroidal fields are written as
\begin{equation}
\frac{\partial A}{\partial t} + \frac{1}{s}(\vec {v_m} \cdot \del)(s A)   = \eta_t\left(\nabla^2 - \frac{1}{s^2}\right)A + \alpha B,
\label{eq:pol}
\end{equation}
\begin{equation}
\frac{\partial B}{\partial t} + \frac{1}{r}\left[\frac{\partial (r v_r B)}{\partial r}+ \frac{\partial (v_\theta B)}{\partial \theta}  \right] = \eta_t\left(\nabla^2 - \frac{1}{s^2}\right)B
+ s(\vec {B_p} \cdot \del){\rm \Omega} + \frac{1}{r}\frac{d\eta _t}{dr}\frac{\partial (rB)}{\partial r},
\label{eq:tor}
\end{equation}
where $A$ is the potential for the poloidal field ($\vec{B_p} = \del \times (A \vec{\hat{\phi}})$, $B$ is the toroidal field,  $s= r\sin{\theta}$, $\vec{v_m} (= v_r \vec{\hat{r}}  +  v_\theta \vec{\hat{\theta}}$) represents the meridional circulation,  $\eta_t$ is the turbulent diffusivity which is assumed to depend only on $r$, $\alpha$ is the $\alpha$ effect, and ${\rm \Omega}$ is the angular frequency.

The term $\alpha B$ in \Eq{eq:pol} is the source for the poloidal field through the $\alpha$ effect. The generation of the poloidal field through the \bl\ process is also parameterised in the 2D (axisymmetric models) through the same $\alpha B$ term. However, this $\alpha$ operates near the surface of the sun and it has a completely different origin than the $\alpha$ effect which operates in the whole convection zone due to helical convection. In comprehensive 3D dynamo models \citep{YM13, MD14,  MT16, Kumar19, BC22}, this $\alpha B$ term is not added in \Eq{eq:pol}, instead, explicit BMRs are deposited whose decay produces a poloidal field.
The source for the toroidal field in \Eq{eq:tor} is due to the ${\rm \Omega}$-effect which is represented by the term: $s(\vec {B_p} \cdot \del){\rm \Omega}$.
The above equations technically represent the equations for the $\alpha{\rm \Omega}$ dynamo model, in which the generation of the toroidal field through the $\alpha$ effect is assumed to be much less than the generation due to $\Omega$ effect, which is true in the sun; see e.g., \citet{CS15}.

\subsection{Causes for long-term variations in the solar activity}

With the above discussion of the solar dynamo, we now identify the causes of the cycle modulation.
As the solar dynamo is nonlinear, it is natural to expect that the modulation in the solar cycle is caused by the back reaction of the flow on the magnetic field.
Therefore, we first identify the nonlinearities in the dynamo models and check if they can lead to cycle modulations.

\subsubsection{Nonlinearities in the dynamo}
As we can see from \Eq{eq:mom}, the magnetic field can alter the flow directly through the Lorentz force.
The Lorentz force can come from the mean magnetic field and the mean current (which is popularly known as the Malkus-Proctor effect \citep{MP75} in the mean-field context) and from the fluctuating magnetic field and the current.
The mean magnetic field can also alter the anisotropic convection which is responsible for transporting angular momentum and maintaining differential rotation and meridional flow in the Sun \citep{Kit94_lambda}.
This effect is also called micro-feedback. When these Lorentz feedbacks of the magnetic fields are included in the flow, we expect a long-term modulation in the flow and the magnetic cycle.

{
In mean-field models, the magnetic feedback is captured by considering a direct Lorentz force of the mean magnetic field in the zonal flow \citep[e.g.,][]{Bushby06} and/or by a quenching term in the $\Lambda$ effect \citep[e.g.,][]{kuker99}.
Cycle modulations in these systems can generally happen in two ways.
In the first one, the magnetic energy of the primary mode (the equatorial symmetry or antisymmetric) can oscillate due to the energy exchange between the flow and the magnetic field via the nonlinear Lorentz feedback.
In this case, a considerable amount of modulation in the differential rotation is observed.
In the second case, a small magnetic perturbation on the differential rotation can slowly change one dominant dynamo mode into another.
In this case, the magnetic field parity can change (between equatorially symmetric (quadrupole) and antisymmetric (dipole)) without producing a large change in the differential rotation.
These two mechanisms are respectively coined as Type II and I modulations.
Mean-field models have demonstrated that nonlinear back reaction of magnetic field on large-scale flow through these types of modulations can induce a variety of modulation patterns in the cycle amplitude, including grand minima and parity modulations which do not leave a strong imprint in differential rotation \citep[e.g.,][]{BTW98, KTW98, Bushby06, WT16}.
Both types of modulation can arise in a model, however, as the observed differential rotation shows a tiny variation over the solar cycle, we expect the Type II modulation is less likely to occur in the Sun.
Even for Type I modulation, a detailed comparison of the magnetic field and the flows in these models with the observations is missing \citep[also see Section 7 of][for a discussion on this topic]{Cha20}.
}


Next is the meridional flow, which
is the second important large-scale flow in the Sun. As it arises due to a slight imbalance between the non-conservative centrifugal and buoyancy forces, we expect its
large variation. In fact, the global simulations
find a large variation in the meridional flow despite a small variation in the differential rotation \citep{Kar15}.
In \bl\ type dynamo models, meridional circulation plays a crucial role in transporting the field on the surface from low to high latitudes and down to the deeper CZ where the shear produces a toroidal field.
The toroidal field is transported to the low latitudes via the equatorward return flow
and possibly causes the equatorward migration
of the sunspot belt. {Thus, in these models, meridional circulation largely} regulates the cycle period \citep{DC99, KC11}. It also affects
the strength of the field as a weak meridional circulation allows the field to advect slowly and gives more time for diffusion \citep{YNM08}. \citet{Kar10} showed that when
a variable meridional flow is used in a high
diffusivity dynamo model to match the observed solar cycle periods, the amplitudes of the cycles are also modelled up to some extent \citep[also see][for modelling various aspects of solar cycle using variable meridional flow]{KC11,Haz15}. In an extreme case, a largely reduced meridional circulation can trigger a Maunder-like grand minimum. In reality, how large the variation in the meridional flow occurred in the past remains uncertain. However, it is obvious that any changes in the flow can lead to modulation in the solar cycle.

Turbulent transport as parameterized by, for example, the
turbulent diffusivity, $\Lambda$ effect, and heat diffusion are also nonlinear because the
Lorentz force of the small-scale as well as
the large-scale dynamo-generated fields act on the small-scale turbulent flows. However, due to limited knowledge in the turbulence theory for solar parameter regions, we do not have a satisfactory model for the magnetic field-dependent form of the turbulent transport parameters; however, see \citet{RK93} and  \citet{KPR94} respectively, for the magnetic field-dependent forms of  $\alpha$ and $\eta$ based on the quasi-linear approximation.

Finally, the toroidal to poloidal part of the dynamo loop involves some nonlinearities. When the generation of poloidal field is due
to the classical $\alpha$ effect, there is a well-known $\alpha$ quenching of the form $ 1 / \left( 1 +  (B/B_{\rm eq})^2 \right) $ with $B_{\rm eq}$ being the equipartition field strength. However, this type of $\alpha$ quenching tries to make a stable cycle rather than producing irregularity in the cycle.
In the \bl\ dynamo, the generation of the poloidal field from the toroidal one also involves
several nonlinearities. Here we discuss the following three potential candidates for these.

\begin{itemize}
\item{Flux loss due to magnetic buoyancy}
\end{itemize}

\begin{figure}
\includegraphics[width=1.0\columnwidth]{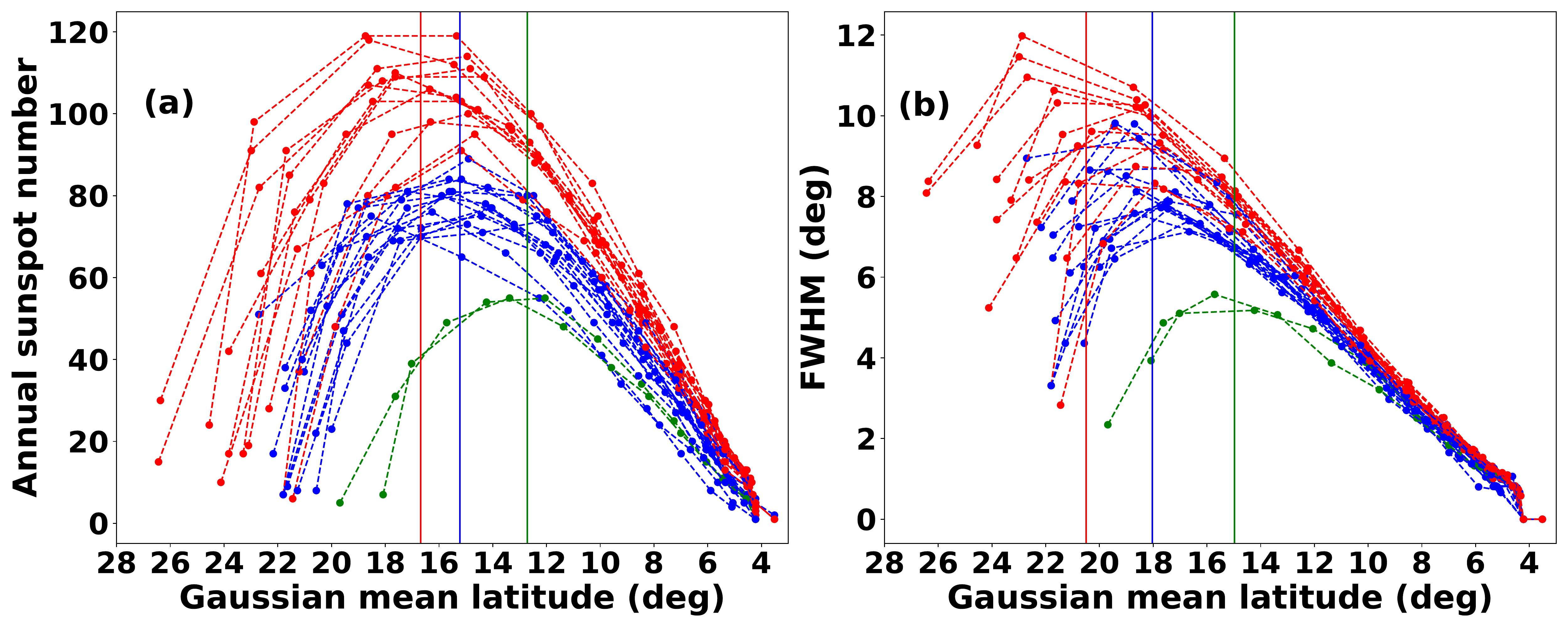}
\caption{The trajectories of (a) annual sunspot number and (b) FWHM vs the central latitude of the
annual spot distribution
obtained from a dynamo simulation with buoyancy-induced flux loss \citep{BKC22}.
Curves clearly show that the beginning phases of the cycles differ widely depending on their strengths but they decline in the same way irrespective of their strengths. This property closely matches with the observations of \citet{CS16}.}
\label{fig:flux_loss}
\end{figure}

The magnetic buoyancy as proposed by \citet{Pa55} plays a critical role in the emergence of BMRs on the solar surface. As the shearing of the poloidal field due to differential rotation intensifies the strength of the toroidal field, there comes a point where the magnetic energy density of the toroidal flux tubes becomes greater than the kinetic energy of the local convective plasma inside the CZ, as a result, the flux tubes become buoyant and start rising through the CZ, eventually giving birth to the sunspots. Following this process, the strength of the magnetic field gets locally reduced as a part of it rises due to buoyancy and the flux tube becomes inefficient to produce further sunspots for some time {\citep[however see a counter-argument by][]{RS01}}.
The sharp rise in the flux loss once the toroidal field strength exceeds a certain value clearly indicates a nonlinear mechanism in the solar dynamo.
Incorporating this mechanism of toroidal flux loss due to buoyancy in a simple manner, \citet{BKC22} showed that this nonlinear process plays a critical role in limiting the growth of the solar dynamo
{which is a potential mechanism to explain why}
different solar cycles rise differently depending on their strength but all the solar cycles decay with similar statistical properties (see \Fig{fig:flux_loss}). They found that introducing the flux loss in the dynamo simulations was critical to reproduce the long-term features of the latitudinal distribution of the sunspots \citep{W55, CS16}; {also see \cite{CS16} and \citet{TNLP22} for an alternative explanation of the universal decay of the solar cycle using cross-equatorial diffusion.}

\begin{itemize}
\item{Latitude quenching}
\end{itemize}

\begin{figure}
\centering
\includegraphics[width=0.6\columnwidth]{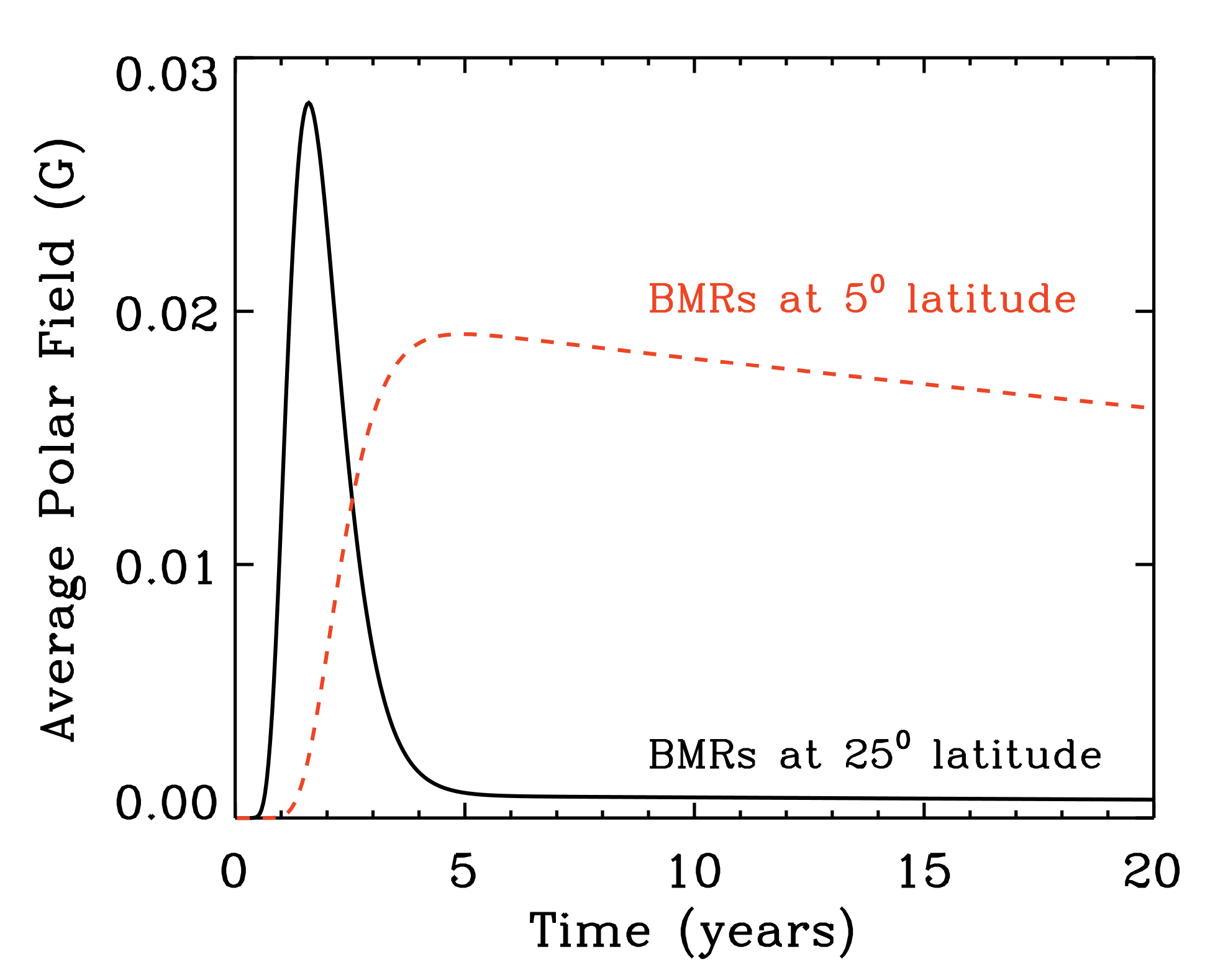}
\caption{
Demonstration of latitude quenching: Temporal evolution of the net polar flux generated from two BMRs deposited symmetrically in two hemispheres at different latitudes.}
\label{fig:lat_quench}
\end{figure}


It has been found that when BMRs appear in low latitudes, the leading polarities from both hemispheres get efficiently cancelled at the equator. 
This leads to the following polarities of the BMRs efficiently getting carried to the poles and {contributing to the polar field}, see \Fig{fig:lat_quench}.
On the other hand, BMRs appearing in the high latitudes do not exhibit efficient cross-hemisphere cancellation and thus do not contribute significantly to the polar field \citep{JCS14,KM18}.
It is seen that strong cycles produce more BMRs at high latitudes. In other words, the average latitude of the BMRs is high for the strong cycles \citep{SWS08, MKB17}.
Hence for a strong cycle, most of its BMRs emerging at high latitudes would be less efficient in polar field production and vice versa for the weak cycles. This mechanism, so-called the {\it{latitude quenching}} \citep{Petrovay20} may help to stabilize the growth of the magnetic field in the Sun \citep{J20}.

Introducing a latitude-dependent threshold on the BMR emergence condition into a 3D \bl\ dynamo simulation, \citet{Kar20} showed that latitude quenching can regulate the growth of a magnetic field when the dynamo is not too supercritical.

\begin{itemize}
 \item{Tilt quenching}
\end{itemize}

\begin{figure}
\centering
\includegraphics[width=0.8\columnwidth]{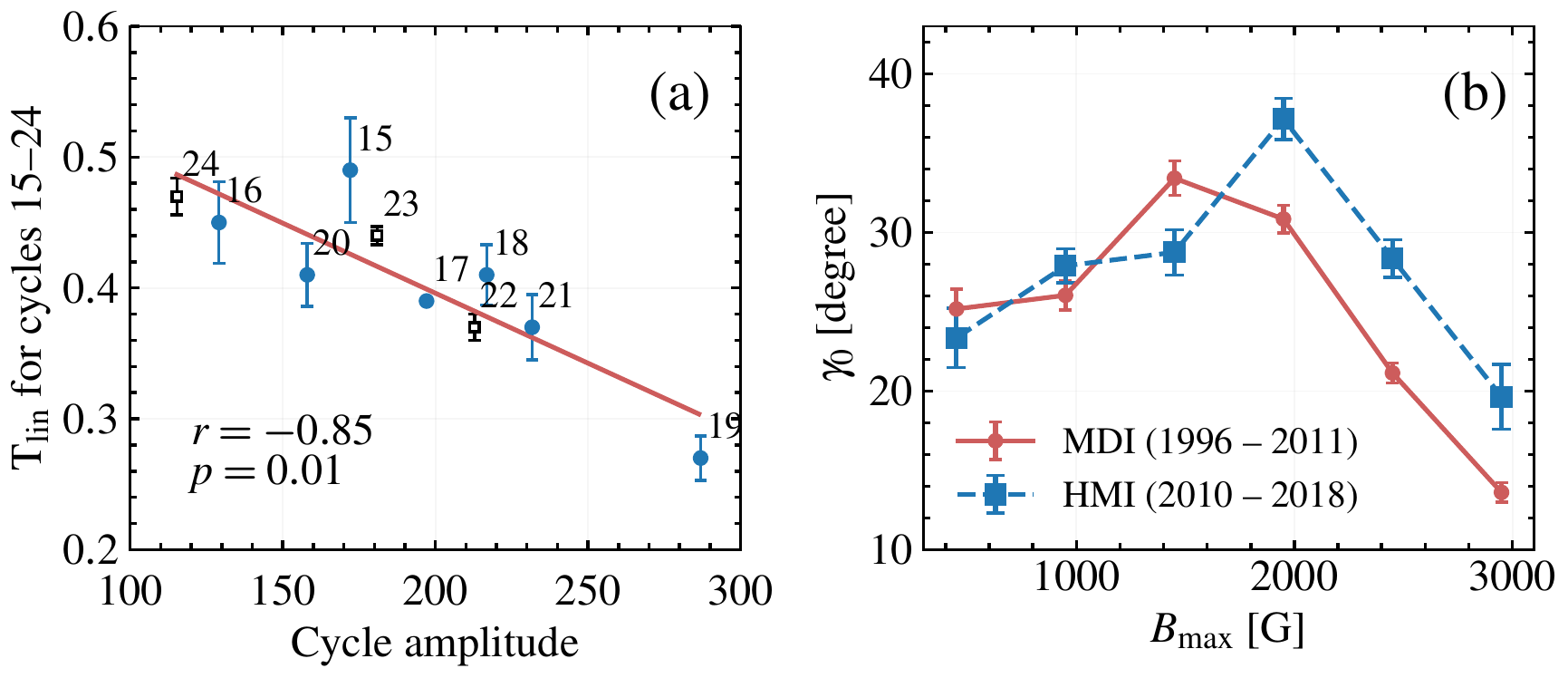}
\caption{Demonstration of tilt quenching: (a) Tilt coefficient (mean tilt normalized by the mean latitude) vs the cycle strength \citep{Jiao21}; also see \citet{Das10}. (b) The slope of Joy's law vs the maximum field strength in the BMR \citep{Jha20}.
}
\label{fig:tilt_quench}
\end{figure}

The tilt angle of BMR plays a crucial role in generating poloidal field in the Sun. For a given latitude, the amount of generated poloidal field increases with the increase of tilt.
The thin flux tube model for the sunspot formation suggests that the tilt of the BMR is produced due to a torque acting on the diverging flows produced from the apex of the rising flux tube which forms the BMR \citep{DC93, FFM94}.
Thus, if the magnetic field of the sunspot-forming flux tube is strong, then it will rise quickly and the Coriolis force will get less time to induce tilt.
In a strong cycle, the toroidal magnetic field is strong and 
{the number of BMRs with strong magnetic field tends to be high \citep{Jha20}.
Thus, we expect the mean tilt in that cycle to be smaller.}
A lesser tilt will produce less poloidal field and the next cycle will be weak. Hence, this may be a potential mechanism for stabilizing the growth of the magnetic cycle through the reduction of tilt which is known as the {\it{tilt quenching}}.

The observational evidence of tilt quenching is limited. \citet{Das10, Jiao21} showed that there is a statistical anti-correlation between the cycle-average tilt of the sunspots with the cycle strength (\Fig{fig:tilt_quench}a). On the other hand, \citet{Jha20} examined the variation of BMR tilt with the strength of its magnetic field within a cycle. They found a non-monotonous dependence of the tilt with the BMR field strength as seen in \Fig{fig:tilt_quench}(b). For weak field strengths, the tilt first increases, however at sufficiently strong field strengths, the BMR tilt starts to decrease.

\subsubsection{Stochastic effects in the dynamo}
The solar convection zone is turbulent and thus the turbulent quantities (such as $\alpha$ effect) are subject to fluctuate around their means. \citet{H93} showed that as there are finite numbers of convection eddies along the longitudes in the sun, the fluctuations of the turbulent transport
coefficients can be larger than their means. There is a long history including the stochastic noise in the $\alpha$ effect in the mean-field dynamo models.  Most of these studies find
long-term modulations in the cycle and grand minima in a certain parameter range of the dynamo number \citep{C92,  OH96, OHS96, GM06, BS08, Mea08}.

\begin{figure}
\centering
\includegraphics[scale=0.50]{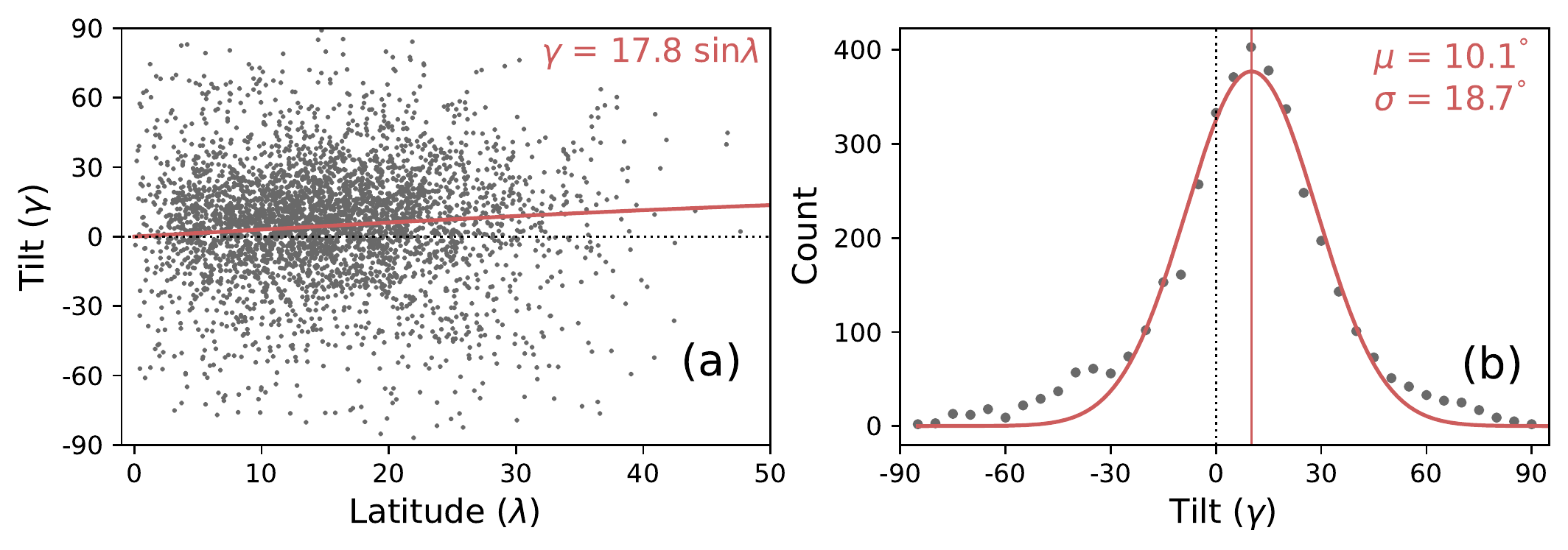}
\caption{(a) Scatter of BMR tilt around Joy's law (solid line). (b) The tilt distribution with fitted Gaussian (solid line).
Here the tilt angles of BMRs are computed by tracking the MDI line-of-sight magnetograms for September 1996 – December 2008.
}
\label{fig:tiltscatter}
\end{figure}

In \bl\ dynamo also
stochastic fluctuations are unavoidable. The toroidal to poloidal part of this model primarily involves stochastic fluctuations due to the following effects.
\begin{itemize}
 \item
 Scatter around Joy's law
\end{itemize}
 Observations find that the tilt ``statistically'' increases with the increase of latitude, which is known as Joy's law. However, a large number of BMRs do not follow this relation (so-called non-Joy), as seen by a huge scatter around the mean trend in \Fig{fig:tiltscatter}. In fact, there are many BMRs which are of anti-Hale type.
These anti-Hale and non-Joy BMRs, having opposite tilts (negative in the northern hemisphere) are responsible for generating opposite polarity field (with respect to the expected polarity) and lead to large fluctuations in the polar field \citep{JCS14, HCM17, Nagy17, Mord22}.

\begin{itemize}
 \item
Variations in the BMR eruption rates
\end{itemize}
There are spatial and temporal variations in the BMR eruptions. BMRs near the equator are much more efficient in generating poloidal field in the Sun because for them the leading polarity can easily connect with the opposite polarity flux from the opposite hemisphere \citep{Ca13, JCS14, KM18, Kar20, Mord22}. Thus variation in the latitudinal position can produce variation in the generated poloidal field.
Next, the rate of BMR eruption is not the same---there is a distribution. Thus, the rate of generation of the poloidal field is not the same \citep{KM17}. Furthermore, the flux contents
of the BMR has also a distribution and thus a wrongly tilted BMR with
{\it high flux}
can disturb the polar field in the sun considerably \citep{Nagy17}.

In summary, the randomness involved in the BMR properties (originated due to the turbulent nature of the convection) produces variation in the poloidal field.
{Although the sun produces thousands of spots in a cycle, only a few spots are produced (on average) per day.
This leads to variations in the polar field comparable to its mean value.}
In the next section, we shall demonstrate some illustrative results from stochastically driven \bl\ dynamo models.

\subsection{\bl\ dynamo models for the long-term variation}
As discussed above, the generation of the poloidal field in the \bl\ dynamo models
involves some randomness. Thus, in axisymmetric dynamo models, these randomnesses
were captured by adding a noise term in the poloidal source \citep[e.g.,][]{DC00}.
Long-term modulations, including Gnevyshev-Ohl/Odd-Even rule \citep{Cha01, Cha07} and grand minima \citep{Cha04, CK09, Passos12, Pas14} are naturally produced in these models. Variations within the cycle, like the amplitude-period anti-correlation \citep{DC00, Kar10} and Waldmeier effect \citep{KC11, BKC22} are also reproduced.
\citet{KMB18} showed that a large variation in the \bl\ process can change the polar field
abruptly and this can lead to double peaks in the following cycle.
While in most of the studies, the level of fluctuations was tuned to produce the observed
variation of the solar cycle including a reasonable number of grand minima,
 \citet{CK12} and \citet{OK13} made some estimate of the fluctuations in the \bl\ process
 from observations.
 \citet{CK12}  found the correct frequency of grand minima as observed in the cosmogenic data for the last 11,000 years.
 \citet{OK13} showed that the statistics of grand minima are consistent with the Poisson random process, indicating the initiation of grand minima to be independent of the history of the past minima.

\begin{figure}
\centering
\includegraphics[width=1.0\columnwidth]{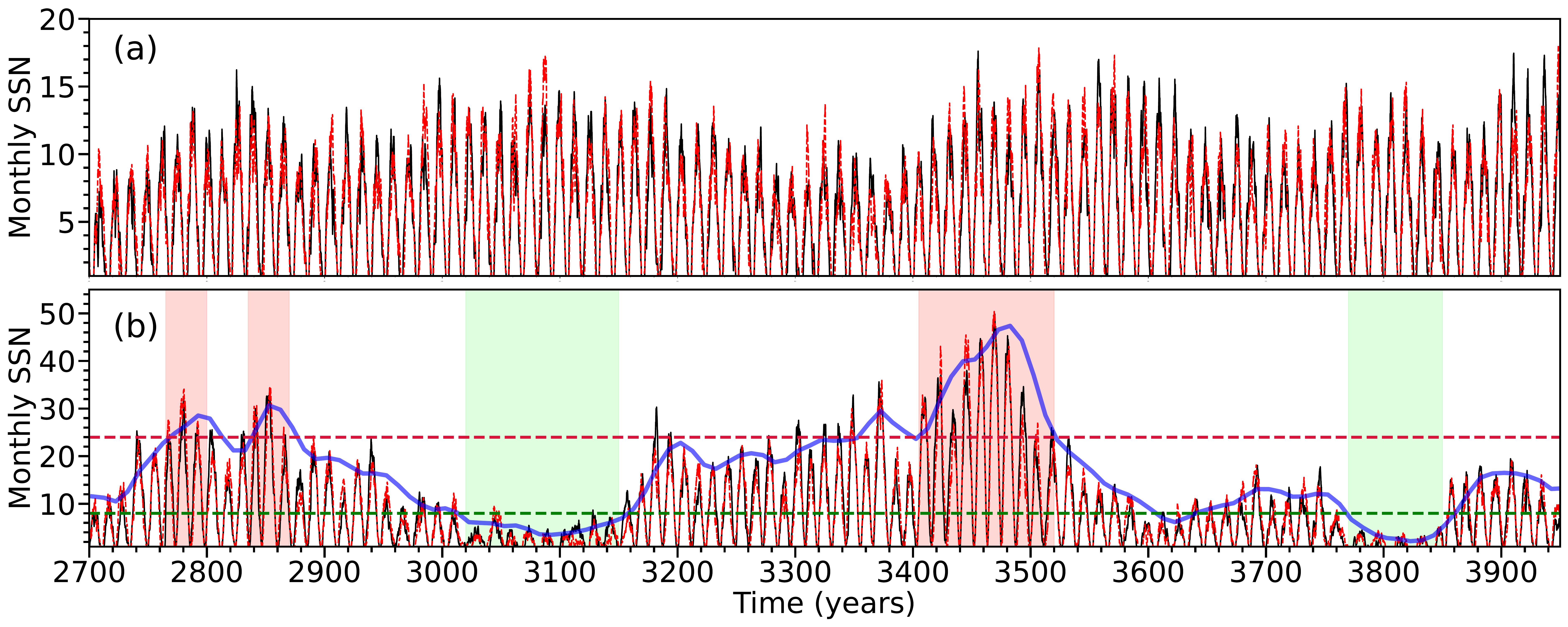}
\caption{Time series of the monthly BMR number
from a 3D dynamo model of \citet{KM17} (a) without tilt scatter around Joy's law and (b) with scatter of $\sigma_\delta=18^\circ$ (close to the observed value). {The black/red curves indicate the north/south hemispheres. The blue curve in panel (b) is the smoothed curve of the cycle trajectories, and the green and red dashed horizontal lines indicate the thresholds for the grand minima and grand maxima, respectively. The green and red shaded regions indicate the grand minima and grand maxima episodes,  respectively. }
}
\label{fig:dyncyc}
\end{figure}

 In recent years, cycle modulations were, {in particular,} produced by including the variations in the BMR properties in two comprehensive models, namely, 2$\times$2D \citep{LC17} and 3D dynamo models \citep{KM17}.
In \Fig{fig:dyncyc}, we show cycles from the 3D dynamo model presented by \citet{KM17}.
As seen in \Fig{fig:dyncyc}(a), the variation in the BMR emergence rate and the flux distribution
produce little variation in the solar cycle.
When the variation around Joy's law tilt is included, it produces a large variation, including suppressed magnetic activity like the one seen during Dalton minimum and Maunder minimum as shown in \Fig{fig:dyncyc}b {{(the regions shaded in green).
Here, the grand minima are identified in the same manner as done in the observed data \citep{usoskin_AA_07}, i.e., the modelled-sunspot data are first binned in 10 years window and smoothed and then a grand minimum is considered when the smoothed data fall below $50\%$ of the average at least for two cycles.
}}

\begin{figure}
\centering
\includegraphics[width=1.0\columnwidth]{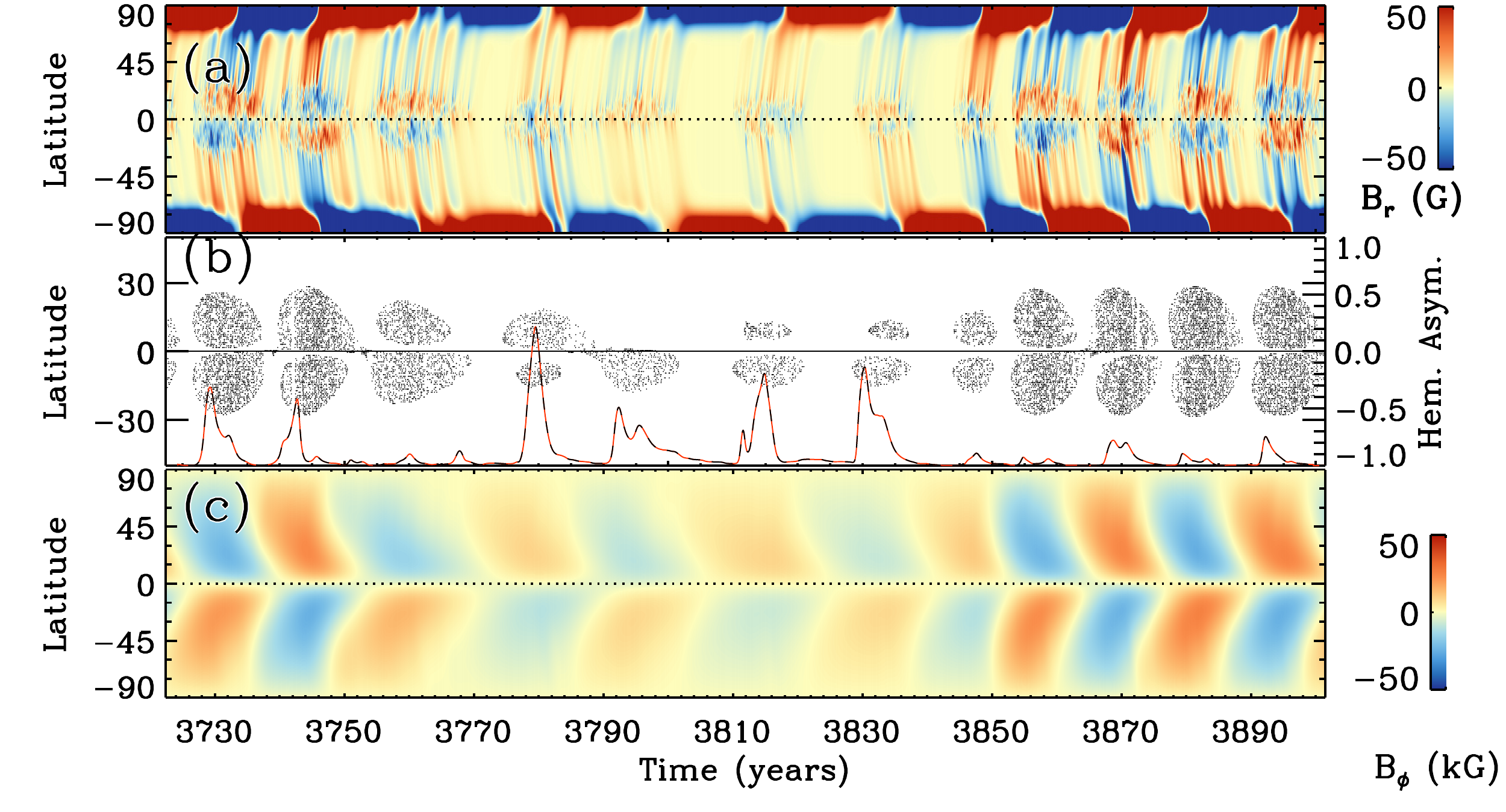}
\caption{Zoomed-in view of a grand minimum presented in \Fig{fig:dyncyc}.
Evolution of  (a) the surface radial field
{{(b) BMR eruptions and hemispheric asymmetry of the toroidal field (black/red curve),}} and (c) the toroidal field at {the bottom of the convection zone}.
}
\label{fig:gm_mod}
\end{figure}

In \Fig{fig:gm_mod}, we present a detailed view of a grand minimum. We find that some of the observed features of the Maunder minimum (hemispheric asymmetry, gradual recovery, slightly longer cycle)
are reproduced in this figure. We note that during this grand minimum, some BMRs are still produced, the number of which is a bit larger than that was observed during \mm\ \citep{Usoskin_15, Vaq15, ZP16,carrasco21}.
However, we should keep in mind that the observations during \mm\ were limited (due to the poor resolving power of the 17th-century telescopes) to detect the small BMRs \citep[e.g.,][]{vaquero09}; only big sunspots were detected.
In our \bl\ dynamo model, few BMRs erupt which produces a poloidal field at a slow rate through the \bl\ process and the model emerges from the grand minimum episode. It is the downward magnetic pumping included in our model which helps to reduce the magnetic flux loss through the surface and recovers the model from grand minima \citep{Ca13, KC16}.

There have been suggestions that during Maunder-like extended grand minima, the \bl\ process may not operate due to few observed sunspots, and $\alpha$ effect \citep{Pa55} is the best candidate for this as it efficiently operates in sub-equipartition field strength \citep{KC13, Pas14, olc19}.
We observe that our model also fails to recover when it enters a deep grand minimum and stops producing BMRs due to the fall of the toroidal field below the threshold for BMR formation. However, this happens very rarely.
While it is a critical question to answer what mechanism dominates in recovering the Sun from an extended grand minimum, it is expected that \bl\ process becomes less efficient during this phase and
the $\alpha$ effect certainly helps in recovering the Sun from grand minima.

Dynamo models with stochastic fluctuations
also produce grand maxima. 
{Our model presented in \Fig{fig:dyncyc}b also produces a few grand maxima shown by the regions shaded in red.
Similar to the grand minima, grand maxima are also computed based on the smoothed sunspot number, but here the threshold is taken as $150\%$ of the long-term mean.}
Systematic studies of grand maxima using dynamo models are limited \citep[however, see][]{KC13, OK13, inceoglu17}.
\citet{KO16} showed that
at the beginning of the cycle, if the generation of the poloidal field is reversed (say due to the emergence of some wrongly tilted BMRs), then it will amplify the existing polar field, instead of reversing it. This increase in the magnetic field can lead to a grand maximum. Another mechanism of grand maxima was given by \citet{olc19}, who showed that when the deep-seated $\alpha$ effect is coupled with the surface \bl\ source, then these
two sources more or less contribute equally to generate a strong poloidal field through a sort of constructive interference.

{Finally, for the secular and supersecular modulations \citep[modulations beyond 11-year periodicity, e.g., Gleissberg cycle, Suess/de Vries cycle, Eddy cycle, and 2400-year Hallstatt cycle;][]{Beer18}, there are limited studies available in the literature.
In a simplified $\alpha$$\Omega$ dynamo model coupled with the angular momentum equation, \citet{Pip99} found the Gleissberg cycle as a result of the re-establishment of differential rotation after the magnetic feedback on the angular momentum transport.
\citet{CS17} modelled the overall power spectrum of solar activity using a generic normal form model for a noisy and weakly nonlinear limit cycle, and \cite{CS19} showed that the long-term modulations beyond the 11-year cycle are consistent with the realization noise, thus casting doubt whether secular and supersecular modulations are connected to the intrinsic periodicities of the solar dynamo.
}

\section{Summary}

Herewith, a brief overview is presented of the long-term variability of solar activity at centennial\,--\,millennial timescales.
The main feature of solar variability is the 11-year quasi-periodic Schwabe cycles, which is however variable \textit{per se} in both magnitudes, duration and phase.
While the direct telescopic observations of the Sun cover roughly four centuries since 1610 and cover a full range of solar-activity levels from the Maunder minimum in the 17th century to the Modern grand maximum in the late 20th century, the quality of the sunspot-number dataset is inhomogeneous and greatly degrades back in time, being quite imprecise before $\approx$1820s.
Moreover, it is too short to study the statistical properties of the solar-cycle modulation on a long timescale.

The cosmogenic-isotope method provides quantitative reconstructions of solar activity on the multi-millennial timescale with stable quality throughout ages making it possible to study long-term solar-cycle modulation.
Using the decadal data for the Holocene (the last twelve millennia), it is possible to identify specific observed properties of solar variability beyond the Schwabe cycle:
\begin{itemize}
\item
The Sun spends about $^1\hspace{-0.07cm}/_6$ of its time in the grand minimum state, grand minima tend to cluster with a $\approx$210-year recurrence time;
\item
The Sun spends about $^1\hspace{-0.07cm}/_{10}$ of its time in the grand maximum state, grand maxima appear without any regular pattern;
\item
During the major fraction of time, the Sun is in the cyclic moderate activity state;
\item
Several quasi-periodicities can be found in long-term solar variability, but they are intermittent and barely significant:
Centennial \textit{Gleissberg} cycle which is an oscillation with the  characteristic time of 60\,--\,140 years;
210-year \textit{Suess/de Vries} cycles manifesting itself through intermittent recurrence of grand minima;
About 2400-year \textit{Hallstatt} cycle whose nature is still unclear; Other long-term cycles, including the millennial Eddy cycle, are insignificant.
\end{itemize}

A recent reconstruction of the annual sunspot numbers from high-precision radiocarbon data for the last millennium makes greatly extended, nearly tripling, the statistic of solar cycles to 96 individually resolved cycles.
In particular, the Waldmeier rule (high cycles rise faster) is statistically confirmed on a larger statistical basis, while the Genvyshev-Ohl rule of the even-odd cycle pairing is not confirmed.
The extended statistic of solar cycles has made it possible, for the first time, to answer the question principle to the solar dynamo theory: is the solar cycle phase-locked, implying an intrinsic synchronisation process as proposed by some external clocking mechanisms, or is random and incoherent.
The new analysis excludes the phase-locking hypothesis at a high significance level, implying that solar cycles vary randomly.

A {brief review} of the theoretical perspectives to explain the observed features in the framework of the dynamo models is presented.
{It is discussed that the nonlinearities in the dynamo, including the effects of the flux loss due to magnetic buoyancy as well as latitude and tilt quenching, help to stabilize the solar dynamo, rather than producing variability in the solar cycle.}
{Primary causes of the solar cycle variability are the stochastic fluctuations in the dynamo which are inherent in different} processes such as a large scatter of the BMR's tilts around Joy's law, and variability in the BMR eruption rates and locations.
On one hand, while modern dynamo models are able to reproduce, with a reasonable ad-hoc tuning of the parameters, the observed features of solar variability, the exact role of those factors is not clear, and some discrepancies between the model results and the data still remain.
On the other hand,
the progress in the
accuracy of models is significant, and we keep gaining knowledge of the processes driving solar variability with the new data acquainted and new models developed.

\bmhead{Acknowledgments}
IU acknowledges the Academy of Finland (project ESPERA No. 321882).
AB and BBK gratefully acknowledge the financial support provided by ISRO/RESPOND (project No. ISRO/RES/2/430/19-20),
the Department of Science and Technology (SERB/DST), India through the Ramanujan Fellowship (project No. SB/S2/RJN-017/2018), the International Space Science Institute (ISSI, Team 474), and the computational resources of the PARAM Shivay Facility under the National Supercomputing Mission, the Government of India, at the Indian Institute of Technology Varanasi.
This work was performed in the framework of the ISSI workshop ``Solar and Stellar Dynamos: A New Era''.

\bmhead{Competing Interests}
The authors declare they have no conflicts of interest.

\bibliography{ISSI_bibtex}


\end{document}